\pdfoutput=1
\documentclass[journal]{IEEEtran}
\usepackage[T1]{fontenc}
\usepackage{textcomp}
\usepackage{lmodern}
\usepackage[latin1]{inputenc}
\usepackage[swedish,english]{babel}
\usepackage{graphics}
\usepackage{amssymb}
\usepackage{amsmath}
\usepackage{graphicx} 
\usepackage{color} 
\usepackage{transparent} 
\usepackage{bm}
\usepackage{fmtcount}
\usepackage{multirow}
\usepackage{caption}
\usepackage{url}
\usepackage{epstopdf}
\usepackage[flushleft]{threeparttable}
\usepackage{babel}
\makeatletter
\adddialect\l@ENGLISH\l@english
\makeatother

\DeclareGraphicsExtensions{.eps}
\newcommand{\bfM}{{\mathbf{M}}}
\newcommand{\bfMp}{{\mathbf{M_\text{p}}}}

\newcommand{\bfH}{{\mathbf{H_{eff}}}}
\newcommand{\dd}{{\mathrm{d}}}

\ifCLASSINFOpdf
\else
\fi

\newcommand{\MYfooter}{\smash{\scriptsize
\hfil\parbox[t][\height][t]{\textwidth}{\centering
~\\
~}\hfil\hbox{}}}

\newcommand{\MYarxivheader}{\smash{\scriptsize
\hfil\parbox[t][\height][t]{\textwidth}{\centering
\copyright\ 2015 IEEE. Personal use of this material is permitted. Permission from IEEE must be obtained for all other uses, in any current or future media, including reprinting/republishing this material for advertising or promotional purposes, creating new collective works, for resale or redistribution to servers or lists, or reuse of any copyrighted component of this work in other works.}\hfil\hbox{}}}

\makeatletter

\def\ps@headings{%
\def\@oddhead{\mbox{}\scriptsize\rightmark \hfil \thepage}
\def\@evenhead{\scriptsize\thepage \hfil \leftmark\mbox{}}
\def\@oddfoot{\MYfooter}%
\def\@evenfoot{\MYfooter}}

\def\ps@IEEEtitlepagestyle{%
\def\@oddhead{\MYarxivheader}%
\def\@evenhead{\scriptsize\thepage \hfil \leftmark\mbox{}}%
\def\@oddfoot{\MYfooter}%
\def\@evenfoot{\MYfooter}}

\makeatother

\begin{document}

\title{Comprehensive and Macrospin-Based Magnetic Tunnel Junction Spin Torque Oscillator Model -- Part I: Analytical Model of the MTJ STO}

\author{Tingsu~Chen,~\IEEEmembership{Student Member,~IEEE,}
        Anders~Eklund,~\IEEEmembership{Student Member,~IEEE,}
        Ezio~Iacocca,~\IEEEmembership{Student Member,~IEEE,}
        Saul~Rodriguez,~\IEEEmembership{Member,~IEEE,}
        Gunnar~Malm,~\IEEEmembership{Senior Member,~IEEE,}
        Johan~$\AA$kerman,~\IEEEmembership{Member,~IEEE,}
        and Ana~Rusu,~\IEEEmembership{Member,~IEEE,} 
\thanks{Manuscript received October 20, 2014; revised December 20, 2014; accepted January 2, 2015. This research is supported by Swedish Research Council (VR).} 
\thanks{Tingsu Chen, Anders Eklund, Saul Rodriguez, Gunnar Malm and Ana Rusu are with the Department
of Integrated Devices and Circuits, KTH Royal Institute of Technology, 164 40 Kista, Sweden.
(e-mail: tingsu@kth.se, ajeklund@kth.se, saul@kth.se, gunta@kth.se and arusu@kth.se).} 
\thanks{Ezio Iacocca and Johan $\AA$kerman are with the Department of Physics, University of Gothenburg, 412 96 Gothenburg, Sweden.(e-mail: ezio.iacocca@physics.gu.se).}
\thanks{Johan $\AA$kerman is also with the Department of Materials and Nano Physics, KTH Royal Institute of Technology, 164 40 Kista, Sweden. (e-mail: akerman1@kth.se).}}

\markboth{IEEE TRANSACTIONS ON ELECTRON DEVICES}{T. Chen \textit{et al.}: Comprehensive MTJ STO model}

\maketitle

\begin{abstract}
Magnetic tunnel junction (MTJ) spin torque oscillators (STO) have shown the potential to be used in a wide range of microwave and sensing applications. 
To evaluate potential uses of MTJ STO technology in various applications, an analytical model that can capture MTJ STO's characteristics, while enabling system- and circuit-level designs, is of great importance.
An analytical model based on macrospin approximation is necessary for these designs since it allows implementation in hardware description languages. 
This paper presents a new macrospin-based, comprehensive and compact MTJ STO model, which can be used for various MTJ STOs to estimate the performance of MTJ STOs together with their application-specific integrated circuits. 
To adequately present the complete model, this paper is divided into two parts. In Part I, the analytical model is introduced and verified by comparing it against measured data of three different MTJ STOs, varying the angle and magnitude of the magnetic field, as well as the DC biasing current. The proposed analytical model is suitable for being implemented in Verilog-A and used for efficient simulations at \mbox{device-}, circuit- and system-levels.
In Part II, the full Verilog-A implementation of the analytical model with accurate phase noise generation is presented and verified by simulations. 

\end{abstract}


\IEEEpeerreviewmaketitle
\section{Introduction}
\IEEEPARstart{T}{he spin} torque oscillator (STO) is a nanoscaled device, utilizing a DC current through a magnetized magnetic multi-layer structure to yield a steady-state voltage oscillation \cite{berger1996emission}-\cite{Kiselev2003}. The operating frequency of this voltage oscillation typically lies in the microwave range \cite{andrei2009} and can be widely tuned by altering the magnetic field and the DC current. 
The typical structure of an STO is presented in Fig.~1(a). 
It is composed of two magnetic layers, the ``free'' layer (FL) and the ``polarizing'' layer (PL, or so-called ``fixed'' layer), being decoupled by a non-magnetic (NM) ``spacer''. 
In this multi-layer structure, a spin-polarized current transfers angular momentum from the PL to the FL, so as to exert a torque on the local magnetization of the FL. By using this transferred torque to compensate the magnetic damping of the FL, magnetization dynamics can be sustained at microwave frequencies. 
The magnetization dynamics of the FL in the presence of spin-polarized current can be described by the Landau-Lifshitz-Gilbert equation with a Slonczewski spin-transfer term (LLGS) \cite{andrei2009}, \cite{slonczewski1999}
\begin{eqnarray}
{\frac{\dd\bfM}{\dd t}} &= -\gamma [\bfM\times\bfH]+\frac{\alpha(\xi)}{M_\text{0}}[\bfM\times{\frac{\dd\bfM}{\dd t}}]+{\gamma}{{\boldsymbol{\tau_\text{STT}}}}
\label{LLGeq}
\end{eqnarray}
where $\gamma$ is the gyromagnetic ratio, $\alpha(\xi)$ is the damping parameter, $M_\text{0}$ is the saturation magnetization, $\bfM$ is the magnetization of the FL, $\bfH$ is the effective magnetic field acting on the FL, and ${{\boldsymbol{\tau_\text{STT}}}}$ is the spin transfer torque (STT), which is used to cancel out the damping term so as to achieve a steady precession of the FL magnetization. ${{\boldsymbol{\tau_\text{STT}}}}$ can be expanded as ${{\boldsymbol{\tau_\text{STT}}}}=a_\text{J}\bfM\times[\bfM\times\bfMp] + b_\text{J}\bfM\times\bfMp$, where $\bfMp$ is the 
magnetization of the PL, $a_\text{J}$ and $b_\text{J}$ are the bias-dependent coefficients of the in-plane and perpendicular torque, respectively. 
The STT has been used in magnetic tunnel junction (MTJ) devices and modeled \cite{STT_model1}, \cite{STT_model2} to develop several novel spintronic devices. 
For example, the STT-based MTJ has been employed in logic circuit designs \cite{logic2012}, \cite{nature2014}, and in the spin transfer torque magnetoresistive random access memory (STT-MRAM) \cite{nature2014}, \cite{augustine2012}, which outperforms SRAM in terms of power consumption and cost, as well as being non-volatile. 
The STT MTJ models \cite{STT_model1}, \cite{STT_model2} used in logic circuits or memory applications, predict the switching between ``0'' and ``1'', then set either the DC resistance or the DC voltage of the MTJ STO accordingly. They do not contain the RF dynamics of STOs and hence are not suitable to present STOs' behavior.
In STOs, the magnetization dynamics are detected by means of the magnetoresistance (MR) effect. This effect predicts a change in the resistance of the multi-layer depending on the relative orientation between the FL and PL. When STT stabilizes a precession of the FL magnetization, an oscillatory resistance is also established leading to an RF voltage generation by virtue of Ohm's law.
This voltage oscillation generated by the STO can be expressed as
\begin{equation}
  V_\text{STO}=R_\text{DC} I_\text{DC}+R_\text{prec}I_\text{DC}\cos(\omega_\text{g}t+\varphi(t))
\end{equation}
where $R_\text{DC}$ is the DC resistance of the STO under a specific biasing condition, $I_\text{DC}$ is the applied current for driving the STO, $R_\text{prec}$ is the amplitude of the resistance oscillation, $\omega_\text{g}$ is the frequency generated by the STO and $\varphi(t)$ represents the random phase fluctuation (i.e. phase noise) of the STO. 

\begin{figure}[tb]
   \centering
  \begin{center}
    \includegraphics[trim = 50mm 75mm 45mm 77mm, clip, width=8.7cm]{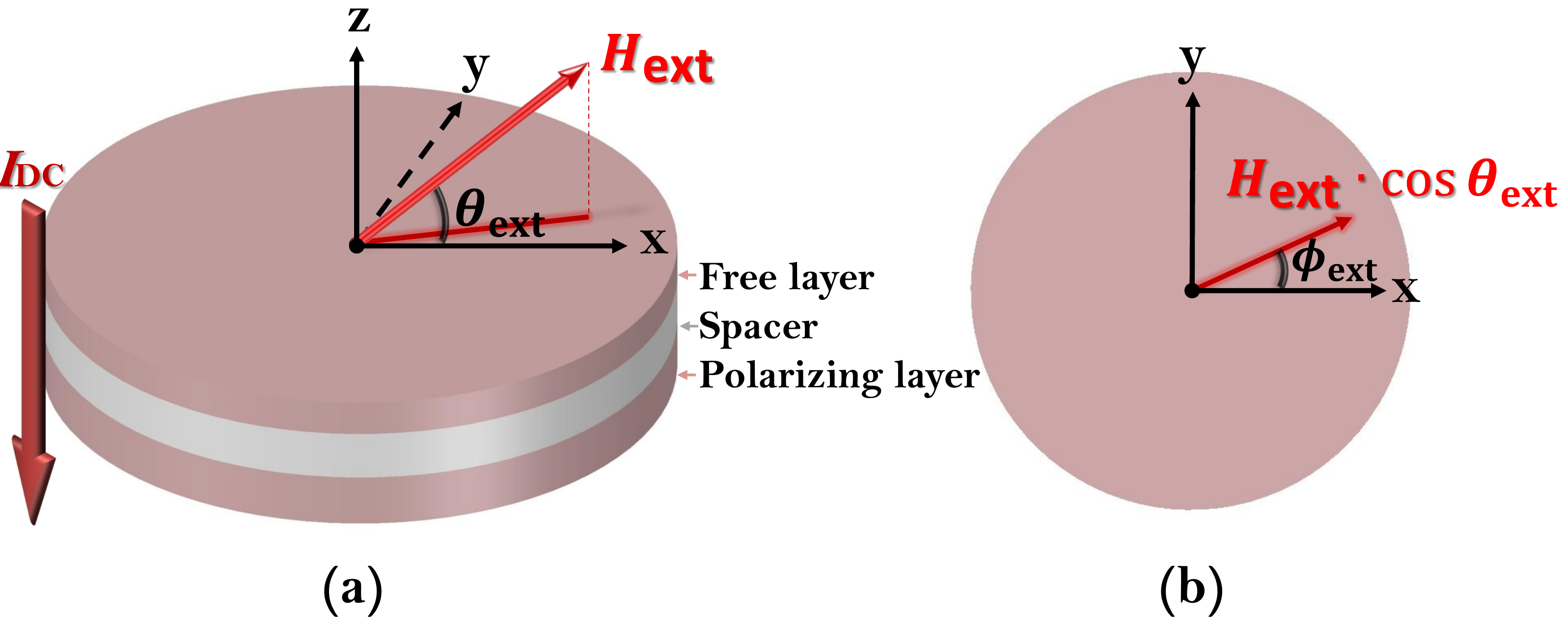}
    \centering
    \caption{Simple structure of the STO showing the directions of the magnetic field and current (a). Normally magnetized by the bias magnetic field $H_\text{ext}$ and bias current $I_\text{DC}$; (b). the in-plane field in an MTJ STO}
    \label{fig-label}
  \end{center}
\end{figure}
STOs exhibit a unique blend of features: high operating frequency, extremely wide tunability, high integration level with CMOS technology, and fast turn-on time (< 1 ns) \cite{Bonetti2009, time1ns}. Thanks to these features, STOs show potential to be used as up-converting mixers without the need for a local oscillator \cite{pufall2005frequency,muduli2010nonlinear}, frequency detectors \cite{nature2014, tulapurkar2005spin}, magnetic field sensors \cite{nature2014,braganca2010nanoscale}, and oscillators \cite{nature2014, Bonetti2009, Tingsu2014_MOTL}. 
Currently, STOs using an insulator as the spacer, in contrast to those using thin metallic spacers, offer higher output power and hence are more suitable for applications. The STO implemented with the thin insulator forms a tunnel junction between the magnetic layers \cite{stiles2006}, and it is the so-called MTJ STO. The MTJ STO is used as the base in this work.

To enable the use of MTJ STOs in applications, a model is necessary to capture their characteristics and to be used in system- and circuit-level designs. 
The core of STO modeling is to solve the LLGS. One possible approach to solve the LLGS is based on micromagnetics \cite{micro_1}, which describe the magnetization dynamics of the STO on a microscopic scale. However, this approach is not possible to implement using a hardware description language, such as Verilog-A, due to the required complicated numeric computation. Another approach is based on the  macrospin approximation, which assumes that only a spatially-uniform magnetization precession is excited and the spin-polarized current is uniform across the area of the free layer \cite{andrei2009}. The macrospin-based STO model is able to provide an analytic solution with acceptable accuracy, while allowing implementation in a hardware description language. As a result, the macrospin-based analytical model can be employed in the design of STO-based systems. 
Two such models, which provide the RF characteristics of the MTJ STO, have been proposed in \cite{MTJSTO_model1, MTJSTO_model2}. 
However, these models contain equations that are usable for matching one specific device and are not fully verified by either experiments or theory. Additionally, the DC operating point of these models has not been analyzed, limiting the accuracy of the model for circuit- or system-level design. Furthermore, only one specific device has been used to verify the models, which is not sufficient.

This paper presents a new comprehensive and compact MTJ STO model, which can overcome the issues of the existing MTJ STO models and be applied to MTJ STOs with arbitrary parameters. Our model is based on the Hamiltonian formalism presented in \cite{andrei2009} as well as extensive analysis of the literature, and it is completed by estimations including but not limited to the DC operating point and electrical RF power. We verify our model by three different MTJ STO measurements under different possible biasing conditions, published by different groups. All the characteristics of the proposed model follow the measurements closely. Furthermore, the proposed analytical model provides a comparable accuracy with the micromagnetics-based model.
Thereafter, it is implemented in Verilog-A, encapsulating the characteristics of MTJ STOs, and is ready for being used by the device and circuit community to implement STO-based systems. This paper is organized as follows. In Part I, the theoretical analyses of the effective magnetic field and the characteristics of the proposed MTJ STO model are provided. Part II describes the implementation of the proposed model in Verilog-A and presents the simulation results of the stand-alone MTJ STO model. Finally, the proposed model is further validated in an STO-based system, where the STO Verilog-A model is simulated together with CMOS RF circuits.

\section{Effective Magnetic Field}
The actual structure of MTJ STOs is complex and can vary in different MTJ STOs \cite{perpendicular_torque3}-\cite{ref3}, as it is given in \mbox{Table I}. However, the generalized stack structure of the MTJ STO, as it is shown in Fig. 1(a) and considered in \cite{andrei2009}, is identical for different MTJ STOs. 
Based on this stack structure, an analytical model of the MTJ STO using the actual cross section size, will be proposed and then verified by comparing it against three different MTJ STOs \cite{perpendicular_torque3}-\cite{ref3} provided by different research groups.
A summary of the important parameters used in this analytical model to obtain the characteristics of these three MTJ STOs \cite{perpendicular_torque3}-\cite{ref3} is presented in Table II. The values of most of the parameters in Table II are available in \cite{perpendicular_torque3}-\cite{ref3} as either measured or suggested values. Most of these values are directly employed in this work. Nevertheless, some values that are provided by \cite{perpendicular_torque3}-\cite{ref3} are slightly adjusted so as to match the analytical model to experiments under different biasing conditions. The adjustments are less than 25$\%$ of the measured or suggested values. Besides, the values of some parameters are not given in \cite{perpendicular_torque3}-\cite{ref3}, as noted in Table II, so that the empirical values from literature are used. $q_1$ and $\eta$ are the only phenomenological parameters and will be detailed later. It should be mentioned that the notations of parameters and STO output characteristics, as well as the notations of external and effective magnetic fields, used in different publications \cite{perpendicular_torque3}-\cite{ref3} are not identical.
To obtain the characteristics of the STO, such as operating frequency, output power, and linewidth (the full width at half-maximum), according to the macrospin-based analytical model in \cite{andrei2009}, the first step is to calculate the effective field. 

\begin{center}
\begin{table}
 \caption{Actual structures of different MTJ STOs}
\label{table:kysymys}
\begin{threeparttable}
\begin{tabular}{lll}
\hline
 & stack of the MTJ STO* & cross section \\ \hline \hline
{\cite{perpendicular_torque3}} & IrMn(5)/CoFe(2.1)/Ru(0.81)/CoFe(1)  & 240$\times$ 240 nm$^2$\\
& /CoFeB(1.5)/MgO(1)/CoFeB(3.5) & \\
{\cite{ref1}} & PtMn(15)/CoFe(2.5)/Ru(0.85) & 140$\times$ 85 nm$^2$\\
& /CoFeB(2.4)/MgO(0.8)/CoFeB(1.8) & \\
{\cite{ref3}} & IrMn/CoFe/Ru/CoFe & 150$\times$ 150 nm$^2$\\ 
& /CoFeB/MgO/CoFeB/NiFe & \\ \hline
\end{tabular}
\begin{tablenotes}
      \scriptsize
      \item *Numbers in parenthesis are the thicknesses in nm. The thicknesses in {\cite{ref3}} are not accessible.
\end{tablenotes}
\end{threeparttable}
\end{table}
\end{center}
\begin{center}
\begin{table}
 \caption{Parameters used for different MTJ STOs}
\label{table:kysymys}
\begin{threeparttable}
\begin{tabular}{lllll}
\hline
 & {defination} & {\cite{perpendicular_torque3}} & {\cite{ref1}} & {\cite{ref3}} \\ \hline \hline
$H_\text{int}$ (Oe) & inter-layer  &   100     &    125*    &  55**      \\
& coupling field &        &        &        \\
$H_\text{A}$ (Oe) & anisotropy field &  100      &  120**      &    5    \\
$M_\text{0}$  & saturation  &    557    &    756    &    515    \\
(emu/cm$^3$) & magnetization &        &        &       \\
$R_\text{AP}$ ($\Omega$) & anti-parallel  &   70     &   650     &      76  \\
 & resistance &        &        &       \\
$R_\text{P}$ ($\Omega$) & parallel resistance &   42.5     &    300    &     45   \\
$\alpha_\text{G}$ & Gilbert damping &    0.12*    &  0.02**      &   0.02**    \\
 & parameter &        &        &        \\
$q_\text{1}$*** & first coefficient in  &    30    &   20    &    25    \\
&$\alpha(\xi)$ expansion &        &        &       \\
$\eta$*** & noise power  &    0.05    &   0.1    &    0.015    \\
$\epsilon$ & dimensionless spin &     0.65**   &  0.7**     &   0.6*     \\
&  polarization efficiency&       &        &       \\ \hline
\end{tabular}
\begin{tablenotes}
      \scriptsize
      \item Values without notes are either measured or suggested values in \cite{perpendicular_torque3}-\cite{ref3}. 
      \item * These values are slightly adjusted based on the measured or suggested values given in \cite{perpendicular_torque3}-\cite{ref3}. 
      \item ** Empirical values.
      \item *** Phenomenological parameters.
\end{tablenotes}
\end{threeparttable}
\end{table}
\end{center}

The effective field, based on \cite{andrei2009}, can be solved by the boundary conditions derived from the integral formulation of Maxwell's equations for magnetic fields. 
However, the boundary conditions (Eq.(102a-102c) in \cite{andrei2009}) take only the magnetic anisotropy field ${H}_{\text{A}}$ into account for simplicity, which is not sufficient to achieve an accurate MTJ STO model since the perpendicular torque and the inter-layer coupling field ${H}_\text{int}$ \cite{perpendicular_torque3}-\cite{jctorque} are considerable in the MTJ STO. The perpendicular torque introduces an additional field with the amplitude of ${b}_{\text{J}}$ (see Eq.1). 
Consequently, the boundary conditions including ${b}_{\text{J}}$ and ${H}_\text{int}$ (${b}_{\text{J}}$ and ${H}_\text{int}$ are both along $\hat{x}$ axis), along $\hat{x}$, $\hat{y}$, $\hat{z}$ axes can be derived as
\begin{subequations}
\begin{eqnarray}
   {H}_\text{eff} \cos\theta_\text{eff} \cos\phi_\text{eff}= {H}_\text{ext} \cos\theta_\text{ext} \cos\phi_\text{ext} \nonumber \\
 + {H}_\text{A} \cos\theta_\text{eff} \cos\phi_\text{eff}-{H}_\text{int}-{b}_\text{J}
\end{eqnarray}
\begin{equation}
 {H}_\text{eff} \cos\theta_\text{eff}  \sin\phi_\text{eff}={H}_\text{ext} \cos\theta_\text{ext} \sin\phi_\text{ext}
\end{equation}
\begin{equation}
  {H}_\text{eff} \sin\theta_\text{eff}={H}_\text{ext} \sin\theta_\text{ext}-4\pi {M}_\text{0} \sin\theta_\text{eff}
\end{equation}
\end{subequations}
where ${H}_\text{eff}$, $\theta_\text{eff}$ and $\phi_\text{eff}$ are the magnitude, out-of-plane and in-plane angles of the effective field; ${H}_\text{ext}$, $\theta_\text{ext}$ and $\phi_\text{ext}$ are the magnitude, out-of-plane and in-plane angles of the external applied field, as illustrated in Fig. 1.
The perpendicular torque coefficient ${b}_\text{J}$, as detailedly studied in \cite{perpendicular_torque3} is a function of the voltage ${V}_\text{DC}$ across the STO.  It is usually determined from the comparison of the theoretical and experimental characteristics. In \cite{perpendicular_torque3}, the expression ${b}_\text{J}$=37$ {V}_\text{DC}$ with the unit of Oe, is used to estimate the additional field introduced by the perpendicular torque for the free-layer excitation mode, where the edge mode is neglected. Although an additional quadratic term has been previously reported \cite{perpendicular_torque2}, its magnitude is relatively small and does not show any significant impact when included in our model. Because of this and the uncertainty in the exact value for this quadratic coefficient, only the linear dependence is implemented. This linear dependence from \cite{perpendicular_torque3} is found to accurately match the characteristics of the MTJ STOs also in \cite{ref1} and \cite{ref3}. 
For the sake of generality, the proposed model allows users changing the coefficients of ${b}_\text{J}$ easily, for the cases where the above expression cannot be successfully applied.

For MTJ STOs, the external field is typically applied in-plane ($\theta_\text{ext}$ = 0) \cite{muduli2010nonlinear, tulapurkar2005spin, Tingsu2014_MOTL, ref1} 
so that the boundary conditions (Eq.(3a-3c)) can be reduced to
\begin{subequations}
\begin{eqnarray}
  H_\text{ext} \sin\phi_\text{ext} \frac{\cos\phi_\text{eff}}{\sin{\phi_\text{eff}}}+H_\text{int}+b_\text{J}=H_\text{A} \cos\phi_\text{eff} \nonumber \\
+H_\text{ext} \cos\phi_\text{ext}
\end{eqnarray}
\begin{equation}
  H_\text{eff}=H_\text{ext} \frac{\sin\phi_\text{ext}}{\sin{\phi_\text{eff}}}
\end{equation}
\end{subequations}
where Eq.(4a) gives the solution to $\phi_\text{eff}$, from which $H_\text{eff}$ can be obtained from Eq.(4b). These simplified equations enable easy implementation of the equation solver in Verilog-A and rapid simulations, so that they are employed to obtain the effective field of MTJ STOs.

\section{Analytical Model of the MTJ STO}
The electrical signal generated by an MTJ STO comprises a DC and an AC (oscillating) component. The DC component of the MTJ STO model can be expressed by either the DC voltage across the MTJ STO or its DC resistance, since the DC current is applied externally. The AC (oscillating) component is characterized by the operating frequency, peak power (or amplitude) and linewidth. 

\subsection{DC operating point} 
In the existing models \cite{MTJSTO_model1, MTJSTO_model2}, the MTJ STOs are assumed to be in the anti-parallel state (with a fixed DC resistance of $R_\text{AP}$), and the DC operating point is not analyzed. 
Nonetheless, in order to compute the other characteristics of the MTJ STO and enable the design of on-chip biasing circuits for it, the DC operating point of the MTJ STO is of importance. The DC voltage across the MTJ STO can be simply calculated by $V_\text{DC}=I_\text{DC}  R_\text{DC}$, where $R_\text{DC}$ is given by \cite{doubling} 
\begin{equation}
  R_\text{DC}=R_\text{P}+(R_\text{AP}-R_\text{P}) \sin^2(\frac{\phi_\text{eff}}{2}) 
\end{equation}
where $R_\text{AP}$ and $R_\text{P}$ are the resistances of the MTJ STO in anti-parallel and parallel states respectively. As it can be noticed from Eq.(5), the DC operating point (or DC resistance) of the MTJ STO can vary greatly as $\phi_\text{eff}$ changes.

\subsection{Operating frequency}
The operating frequency $\omega_\text{g}$, according to \cite{andrei2009}, is given by
\begin{equation}
  \omega_\text{g}=\omega_\text{0}+N \overline{p}
\end{equation}
where $\omega_\text{0}$ and $N$ are the ferromagnetic resonance (FMR) frequency and coefficient of the nonlinear frequency shift respectively, which can be obtained from Eq.(103), Eq.(104a) and Eq.(105a) in \cite{andrei2009} once the effective field is determined; $\overline{p}$ is the dimensionless power, which is a coefficient that is proportional to the experimentally measured RF power of the STO \cite{andrei2009} and is involved in calculating all the characteristics of MTJ STOs, hence it is a critical quantity. 
As detailed in \cite{andrei2009} (Eq.(84b)), $\overline{p}$ is a function of the supercriticality $\zeta$, nonlinear damping coefficient $Q$ and noise power $\eta$. $\zeta$ is defined as the ratio between $I_\text{DC}$ and the threshold current $I_\text{th}$ of the MTJ STO. $I_\text{th}$ (see Eq.(22) in \cite{andrei2009}) is a function of $\phi_\text{eff}$, the nonlinear damping rate $\Gamma_\text{G}$ calculated based on the effective field (see Eq.(104b) in \cite{andrei2009}), and the coefficient $\sigma_\text{0}$ expressed as \cite{andrei2009}
\begin{equation}
  \sigma_0=\frac{\epsilon g \mu_\text{B}}{2  e   M_\text{0}  l   A} 
\end{equation}
where $\epsilon$ is the dimensionless spin polarization efficiency, $g$ is the spectroscopic Land$\acute{\text{e}}$ factor, $\mu_\text{B}$ is the Bohr magneton, $e$ is the modulus of the electron charge, $l$ is the thickness of FL, and $A$ is the area of the current-carrying region. For MTJ STOs with CoFe and CoFeB as the PL, $\epsilon$ is $0.65$ $\pm$ $0.05$ and $0.56$ $\pm$ $0.03$, respectively \cite{epsilon1,epsilon2}.  
To obtain the other important parameter $Q$ that is required to solve $\overline{p}$, $q_1$ involved in the expression of $Q$ (see Eq.(105b)) needs to be determined. $q_1$ is the first coefficient in the expansion of $\alpha(\xi)$ and is usually considered as a phenomenological parameter \cite{andrei2009}. The condition for $q_1$ is that the resulting $Q$ falls within its typical range $0\le Q\le3$ \cite{andrei2009}. Likewise, the noise power $\eta$ is typically treated as a constant value  between 0 and  0.2 \cite{andrei2009}. To determine $q_1$ and $\eta$, they need to be roughly adjusted and tested together in order to enable a good agreement of the MTJ STO characteristics between the proposed model and the experimental results. 

To validate the operating frequencies of the proposed MTJ STO model, we compare the numerical results to the experimental measurements provided by \cite{perpendicular_torque3}-\cite{ref3} under different biasing conditions, and the comparison is depicted in Fig. 2. 
\begin{figure}[tb]
   \centering
  \begin{center}
    \includegraphics[trim = 70mm 71.4mm 70mm 50.5mm, clip, width=7.7cm]{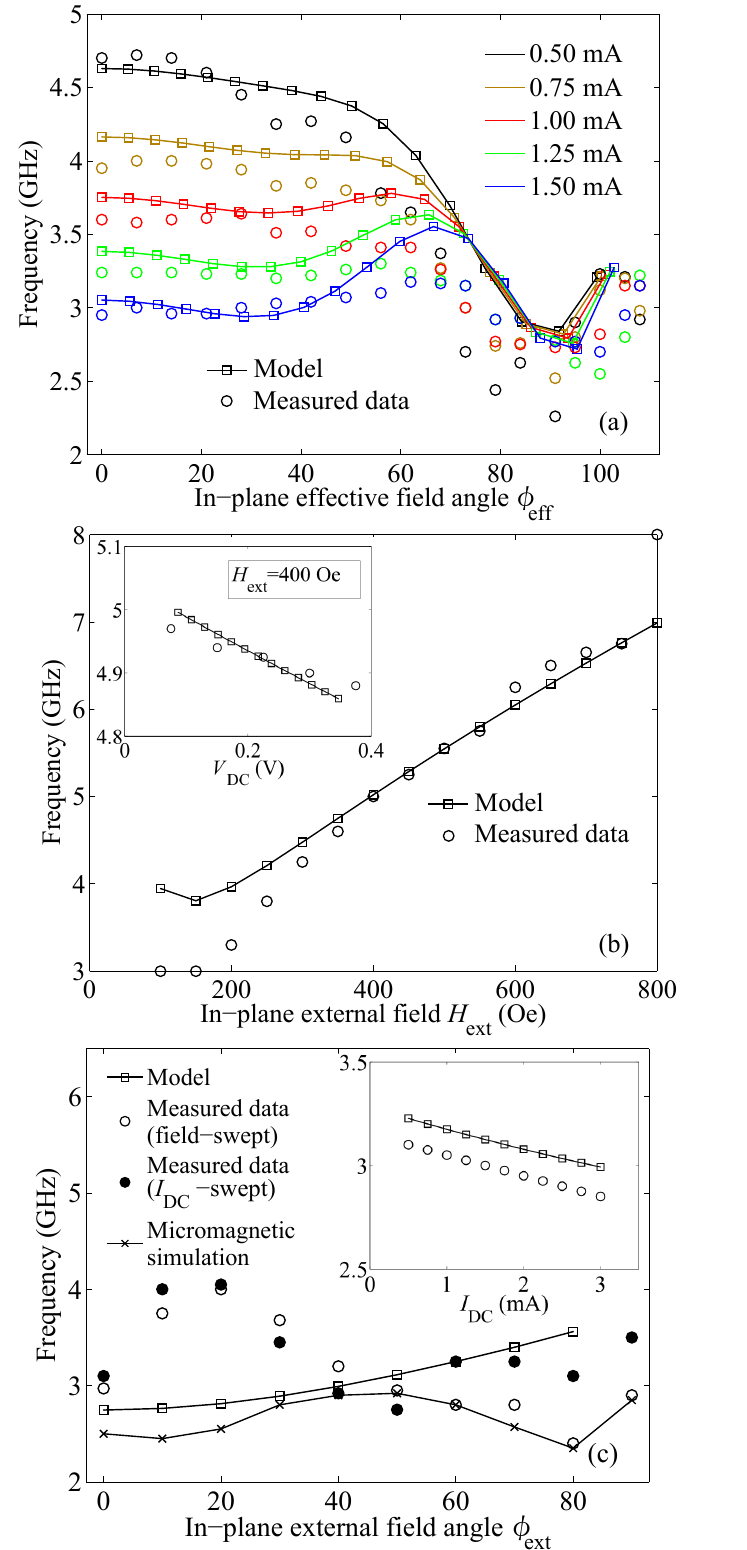} 
    \centering
    \caption{Comparison of simulated $\omega_\text{g}$ of the proposed model with measured $\omega_\text{g}$ extracted from (a). \cite{ref1} (b). \cite{perpendicular_torque3} and (c). \cite{ref3}}
    \label{fig-label}
  \end{center}
\end{figure}
Fig. 2(a) shows the comparison between the model and the measured $\omega_\text{g}$ (\cite{ref1}) at $H_\text{ext}$ = 300 Oe, with respect to $\phi_\text{ext}$ as well as $I_\text{DC}$. 
For $\phi_\text{eff}$ < 70$^o$, as $I_\text{DC}$ increases, a decreasing $\omega_\text{g}$ is observed in both the modeled and the measured results as presented in Fig. 2(a).
At $\phi_\text{eff}$ = 70$^o$ and 90$^o$, $\omega_\text{g}$ are similar for different $I_\text{DC}$. This is mainly because the nonlinear frequency shift $N$ in Eq.(6) is close to zero at these angles, so that the operating frequencies for different $I_\text{DC}$ are mostly determined by the FMR. This observation is in agreement with the measured $N$ \cite{ref1}. 
Generally, the operating frequency of the proposed model matches the measured data in a large region. The discrepancy between the model and the measured data may be due to the simplification of the macrospin-based model, which is not able to represent the complexities of a real device such as microscopic dynamics, roughness and the coupling between layers.

To further verify the dependence of the external magnetic field on $\omega_\text{g}$ of the proposed model, $\omega_\text{g}$ of the model as a function of the external field and the biasing voltage $V_\text{DC}$, is compared with the measured $\omega_\text{g}$ given in \cite{perpendicular_torque3}, and illustrated in Fig. 2(b). The results obtained from the model matches well with the measured data.
For low magnetic fields, a small discrepancy between the model and measured results can be found. This could be due to the fact that the layers are not fully saturated under low magnetic fields. 

The measured $\omega_\text{g}$ in \cite{ref3} and $\omega_\text{g}$ obtained from the micromagnetics simulation results \cite{ref3}, are also used to compare with $\omega_\text{g}$ of the model. In \cite{ref3}, the MTJ STO was measured under a fixed external field of 200~Oe, so that the comparison is made under this condition. 
The comparison is shown in Fig. 2(c). 
Within 0$^o$ < $\phi_\text{ext}$ < 50$^o$, the proposed model offers good agreement with the results obtained from the micromagnetic-based model. However, similar discrepancies are identified between the measured results and their respective models. This may be caused by the multiple modes of the sample, where the frequencies of the first and second modes appear close to each other at low field magnitudes and low field angles \cite{ref3}. The multi-mode behavior is neither captured by the macrospin-based model, nor correctly estimated by the micromagnetics-based model. Regarding the large field angles (50$^o$ < $\phi_\text{ext}$ < 90$^o$), inconsistency of measured results is found between $I_\text{DC}$- and field-swept measurements. This might be explained by electromigration, which degrades the MTJ STO during the measurement. In this case, no model can completely predict the operation of the MTJ STO; yet, both of the macrospin- and micromagnetics-based models provide reasonable agreement with $I_\text{DC}$- and field-swept measurements respectively.
The comparison in Fig. 2(c) indicates that the proposed analytical model is able to capture the core behaviors of the MTJ STO in spite of the considerable complexities in the MTJ STO, which are not yet totally identified and fully understood. In addition, the model provides the operating frequencies of MTJ STOs, which have similar accuracy as the ones obtained from the micromagnetic simulations. 
Moreover, the operating frequencies measured by sweeping $I_\text{DC}$ at a fixed external field angle $\phi_\text{ext}$ = 40$^o$ that are available in \cite{ref3}, are compared with the modeled results. For different $I_\text{DC}$, the proposed model matches the measured data with a slight frequency offset of about 0.15 GHz (Fig. 2(c)), which is less than 5\% of the operating frequency.

\subsection{Peak power}
For estimating the peak power generated by the MTJ STO, expressions determining the magnetoresistance and precession angle were introduced and used in the existing MTJ STO models \cite{MTJSTO_model1, MTJSTO_model2}. These expressions, however, are not yet validated by either the theory or experiments. In this work, the expressions that have been typically employed to estimate the precession angle $\theta_\text{prec}$ based on the measured power \cite{time1ns, perpendicular_torque2} and validated by experiments \cite{doubling}, are used. The peak power of the fundamental signal of the MTJ STO is given by \cite{perpendicular_torque2} 
\begin{equation}
  P(\omega)=\xi(\omega) (\frac{R_\text{AP}-R_\text{P}}{R_0})^2  J_1^2(\theta_\text{prec}) \sin^2\phi_\text{eff}   \frac{R_0 I_\text{DC}^2}{8}
\end{equation}
where $\xi(\omega)$ is the RF power transfer efficiency at frequency $\omega$ and assumed to be 1 (no loss due to parasitics), 
$J_1(\theta_\text{prec})$ is the Bessel function of the first kind, and $R_0$ is defined as  
\begin{equation}
  R_0=\frac{R_\text{AP}+R_\text{P}}{2}-\frac{R_\text{AP}-R_\text{P}}{2}  J_0(\theta_\text{prec})  \cos\phi_\text{eff} 
\end{equation}
To model the output power of the MTJ STO, an expression for estimating $\theta_\text{prec}$ is required to determine the peak power generated by the MTJ STO using Eq.(8).
$\theta_\text{prec}$ can be estimated based on the equation provided by \cite{pp2009}, which is re-written as 
\begin{equation}
  \theta_\text{prec}=2 \arcsin(\sqrt{\overline{p}})
\end{equation}
Eq.(10) gives the dependence of $\theta_\text{prec}$ on the dimensionless power $\overline{p}$ converted from the experimental results. Here, the dimensionless power $\overline{p}$ has already been obtained in the previous subsection based on the analytical theory.
 
Eq.(8) and Eq.(9) contain Bessel functions of the first kind, which are difficult to implement in Verilog-A. 
Since MTJ STOs have limited precession angles \cite{doubling}, based on the expansions of Bessel functions, two approximations, including $J_1(\theta_\text{prec})\sim \frac{\theta_\text{prec}}{2}$ and $J_0(\theta_\text{prec})\sim 1-\frac{\theta_\text{prec}^2}{2^2}$, can be used to simplify Eq.(8) and Eq.(9). 
In addition, in order to reproduce the voltage oscillation as described by Eq.(2) in the Verilog-A model, $R_\text{prec}$ in Eq.(2) can be derived and simplified based on Eq.(8) as 
\begin{equation}
  R_\text{prec}=(\frac{R_\text{AP}-R_\text{P}}{R_0})^2 \theta_\text{prec}^2  \sin^2\phi_\text{eff}  \frac{R_0}{32}
\end{equation}
Eq.(9) can be also reduced to
\begin{equation}
  R_0=\frac{R_\text{AP}+R_\text{P}}{2}-\frac{R_\text{AP}-R_\text{P}}{2}  (1-\frac{\theta_\text{prec}^2}{2^2})  \cos\phi_\text{eff}
\end{equation}
The peak power and the voltage amplitude of the fundamental signal of the MTJ STO can be determined based on Eq.(11) and Eq.(12). The power generated by the second harmonic is not considered in this work, yet can be computed in a similar way and added to the proposed model. 

Figure 3 presents the comparison of the peak power (a) and the peak power spectral density (PSD) (b) between the proposed model and the available experimental results (\cite{ref1,ref3}). 
\begin{figure}[tb]
   \centering
  \begin{center}
    \includegraphics[trim = 23mm 96.9mm 120mm 98.4mm, clip, width=7.1cm]{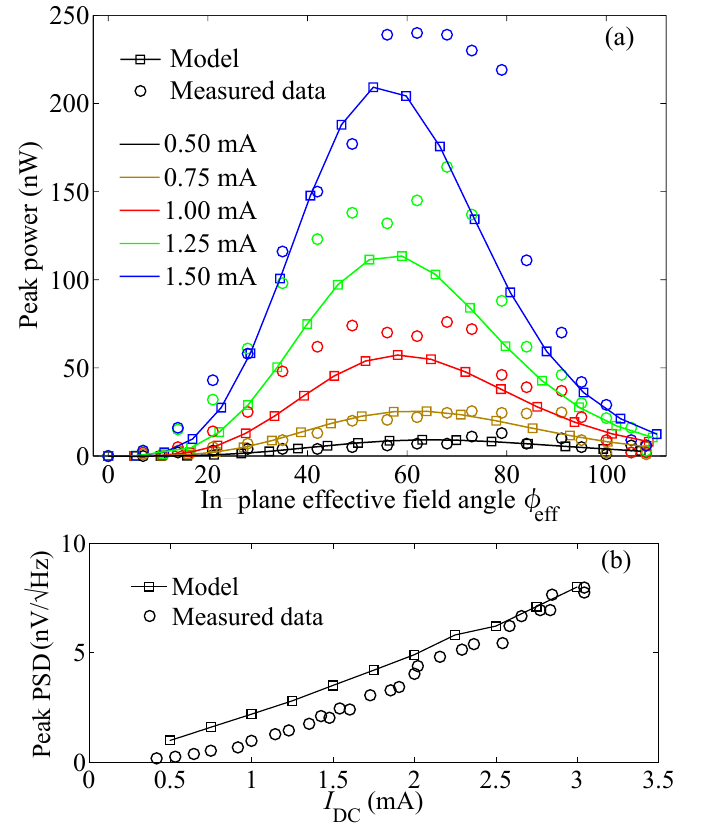}
    \centering
    \caption{The calculated peak power and peak PSD of the proposed model are compared with the measured data taken from (a). \cite{ref1} and (b). \cite{ref3}}
    \label{fig-label}
  \end{center}
\end{figure}
The output peak power of the proposed model, presented in Fig. 3(a), is derived as $(R_\text{prec}  {I_\text{DC}})^2 / R_\text{DC}$ and compared with the measured peak power \cite{ref1}.
It follows the measured data \cite{ref1} under different biasing conditions, where the $I_\text{DC}$ and $\phi_\text{eff}$ are swept. The minor discrepancies between the modeled and measured peak power might be due to the imperfections of loss de-embedding during the measurement.

To further validate the peak power (or voltage amplitude) of the proposed model as a function of $I_\text{DC}$, the time domain signals generated by the proposed model are converted to the frequency domain to obtain their PSDs, so as to compare the peak PSDs against the measured ones in \cite{ref3}. 
To generate these signals for evaluating the amplitudes of the spectrum, the required information of linewidth is taken from the measured data \cite{ref3}.
In addition, as it can be calculated from \cite{ref3}, the sum of peak PSDs of the second and third modes takes approximately 1/3 of the total peak PSDs of all the modes. Accordingly, a coefficient of 0.67 is used to downscale the modeled peak PSDs, in order to compare the modeled peak PSDs of the first mode with the measured ones when multiple modes exist.
The comparison of the modeled and measured peak PSDs is shown in Fig. 3(b). 
For different $I_\text{DC}$, the peak PSDs obtained from the proposed MTJ STO model are very close to those given by the experiments \cite{ref3}.

\subsection{Linewidth}
In the existing MTJ STO models \cite{MTJSTO_model1, MTJSTO_model2}, the equation used to obtain the linewidth $2 \Delta \omega$ is given by Eq.(95) in \cite{andrei2009}
\begin{equation}
  2 \Delta \omega=(1+\nu_\text{fs}^2)\Gamma_\text{+}(p_\text{0}) \frac{k_\text{B}T}{\varepsilon(p_\text{0})}
\end{equation}
where $\nu_\text{fs}$ is the normalized dimensionless nonlinear frequency shift, $\Gamma_\text{+}(p_\text{0})$ is the positive damping rate, $k_\text{B}T$ is the product of the Boltzmann constant and the temperature, and $\varepsilon(p_\text{0})$ is the oscillation energy. $\nu_\text{fs}$, $\Gamma_\text{+}(p_\text{0})$ and $\varepsilon(p_\text{0})$ can be computed based on the known parameters and the solved nonlinear coefficients (see Eq.(33), Eq.(19b) and Eq.(77) in \cite{andrei2009}).
Nonetheless, Eq.(13) is only valid for the above-threshold regime, where $I_\text{DC}$ is larger than $I_\text{th}$ \cite{andrei2009}. 
Thus, it should not be applied to all the biasing conditions. Regarding the below-threshold regime ($I_\text{DC}<I_\text{th}$), $2\Delta \omega$ is found to be \cite{andrei2009}
\begin{equation}
  2 \Delta \omega=2\Gamma_\text{G}(1-\frac{I_\text{DC}}{I_\text{th}})
\end{equation}
Noticed from Eq.(14), as the MTJ STO operates in the near-threshold regime ($I_\text{DC}\approx I_\text{th}$), $2 \Delta \omega$ goes to zero, which cannot be achieved in practical cases. 
Our calculations suggest that Eq.(14) is not valid for the regime $0.85  I_\text{th}<I_\text{DC}\le I_\text{th}$, since the computed linewidth is much lower than practical values. Hence, the regime $0.85  I_\text{th}<I_\text{DC}\le I_\text{th}$ is considered as the near-threshold regime in this work.
However, as explained in \cite{andrei2009}, the linewidth of an MTJ STO operating in near-threshold regime, is difficult to approximate theoretically. 
Fortunately, our calculations show that in this near-threshold regime, using Eq.(13) can offer reasonable agreement between the modeled and the measured linewidth. 
Since this work is targeting MTJ STO modeling rather than exploring the linewidth of the MTJ STO, Eq.(13) is simply applied to approximate the linewidth in the near-threshold regime, even though, it is not theoretically validated.
\begin{figure}[tb]
   \centering
  \begin{center}
    \includegraphics[trim = 23mm 97mm 120mm 98.4mm, clip, width=7.1cm]{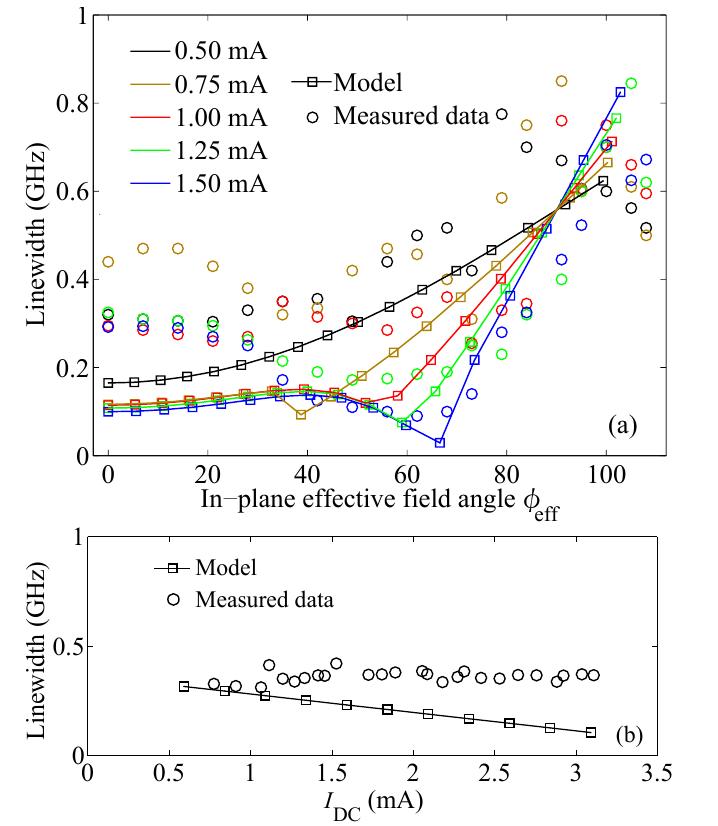}
    \centering
    \caption{The calculated $2\Delta \omega$ of the proposed model is compared with the measured $2\Delta \omega$ taken from (a). \cite{ref1} and (b). \cite{ref3}}
    \label{fig-label}
  \end{center}
\end{figure}

To verify the linewidth of the proposed model, the available measurement results of two different MTJ STOs \cite{ref1, ref3} are employed. 
Figure 4(a) shows the comparison between the measured $2 \Delta \omega$ given in \cite{ref1} and the $2 \Delta \omega$ obtained from the proposed model, as a function of $\phi_\text{ext}$ and $I_\text{DC}$. For different angles, the general trend of the modeled $2 \Delta \omega$ follows the trend of the measured $2 \Delta \omega$, owing to the different equations used for computing the linewidth in different regimes.
Besides, for low currents ($I_\text{DC}$ = 0.5 mA), the linewidth of the proposed model shows an increasing trend as a function of the field angle, indicating that the MTJ STO is operating in the below-threshold regime. The threshold point shifts towards higher field angles as $I_\text{DC}$ increases.
Figure 4(a) demonstrates that the proposed model is able to offer the linewidth with the correct order of magnitude and proper trend regardless of the biasing conditions, despite the huge simplification of linewidth estimation provided by the macrospin approach.

The linewidth of the proposed model is also compared with the measured data provided in \cite{ref3}, and depicted in Fig. 4(b). As $I_\text{DC}$ increases, the measured $2 \Delta \omega$ remains almost flat, while the $2 \Delta \omega$ of the model is decreasing. The trend of the linewidth obtained from the model, yet, agrees with the theoretical analysis in \cite{andrei2009}. 
The disagreement is likely due to the fact that current induces stronger dynamics, which physically translates to the generation of higher order spin wave modes and possibly mode-hopping events that broaden the linewidth \cite{iacocca2014}. For MTJ STOs with sufficiently large cross section, such modes can only be represented micromagnetically, affecting the accuracy of the macrospin approach.
Moreover, MTJ STOs have been reported to exhibit not only the analytically treated white frequency noise but also colored, $1/f$ type noise for low fluctuation frequencies \cite{quinsat2010}. Frequency noise of the $1/f$ type has been shown to lead to an increased measured linewidth \cite{keller2010}, making it another plausible explanation of the discrepancy between the measured data and the proposed model. 

\section{Conclusion} 
A comprehensive and compact analytical MTJ STO model based on the macrospin approximation as well as the physics-based equations of the STO, has been proposed. 
The perpendicular torque and inter-layer coupling field that significantly affect the characteristics, have both been considered in the proposed model to calculate all the characteristics of MTJ STOs. 
The model has been compared under different biasing conditions against the experimental data obtained from three different MTJ STOs. 
Despite its simplicity, the model can reproduce the experimental data with an acceptable degree of accuracy.  
Therefore, it is suitable for being implemented in a hardware description language, which enables the evaluation and utilization of MTJ STOs in real and extensive applications.

\ifCLASSOPTIONcaptionsoff
  \newpage
\fi

\bibliographystyle{plain}

\begin{IEEEbiography}[{\includegraphics[trim = 34mm 82mm 38mm 44mm width=1in,height=1.25in,clip,keepaspectratio]{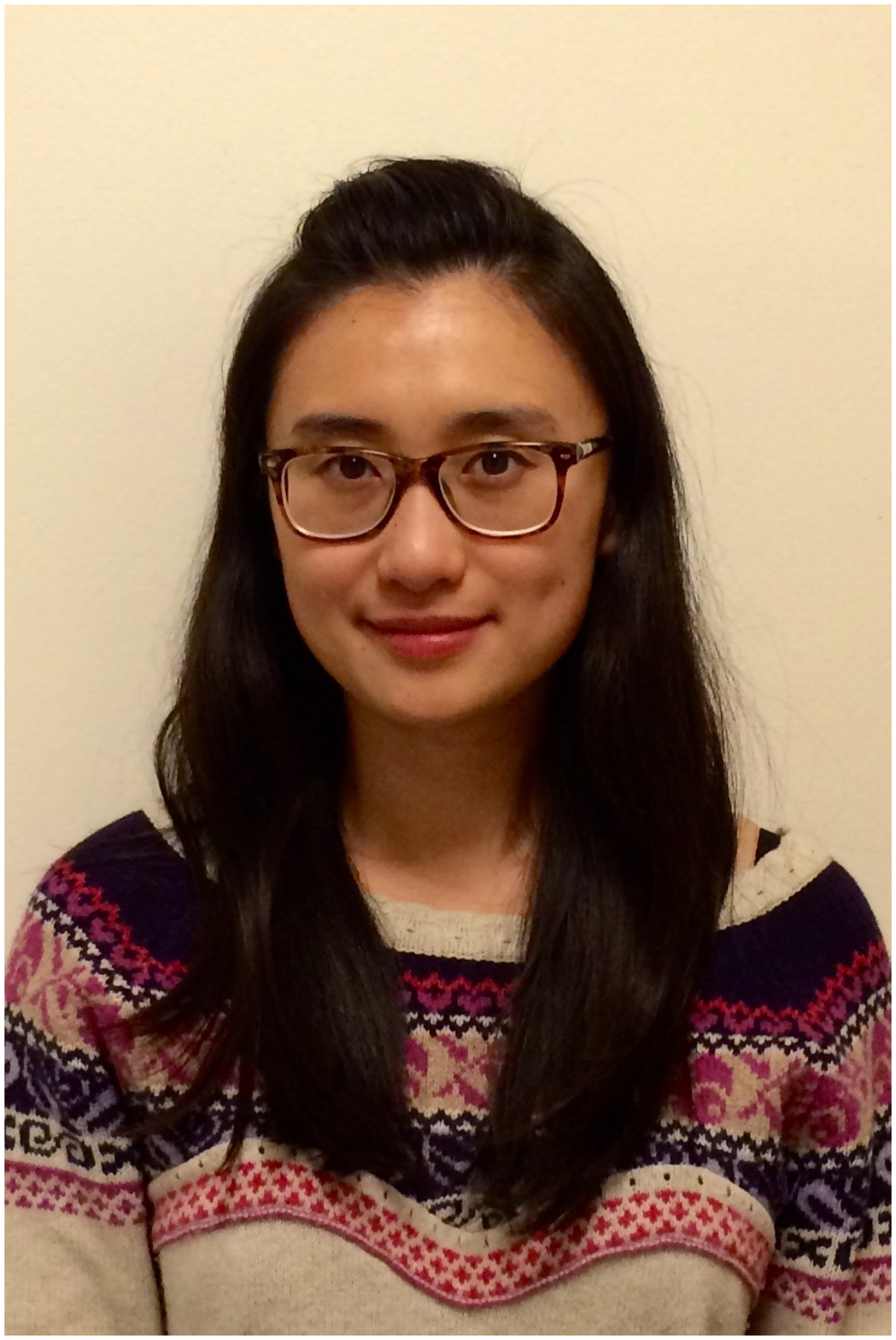}}]
{Tingsu Chen}
(S'11) received the B.Sc. degree in communication engineering from the Nanjing University of Information Science and Technology, China, and the M.Sc. degree in system-on-chip design from the KTH Royal Institute of Technology, Sweden, in 2009 and 2011, respectively. She is currently working toward the Ph.D. degree at KTH with the research area of high frequency circuit design for spin torque oscillator technology.
\end{IEEEbiography}
\begin{IEEEbiography}[{\includegraphics[trim = 37mm 88mm 62mm 44mm width=1in,height=1.25in,clip,keepaspectratio]{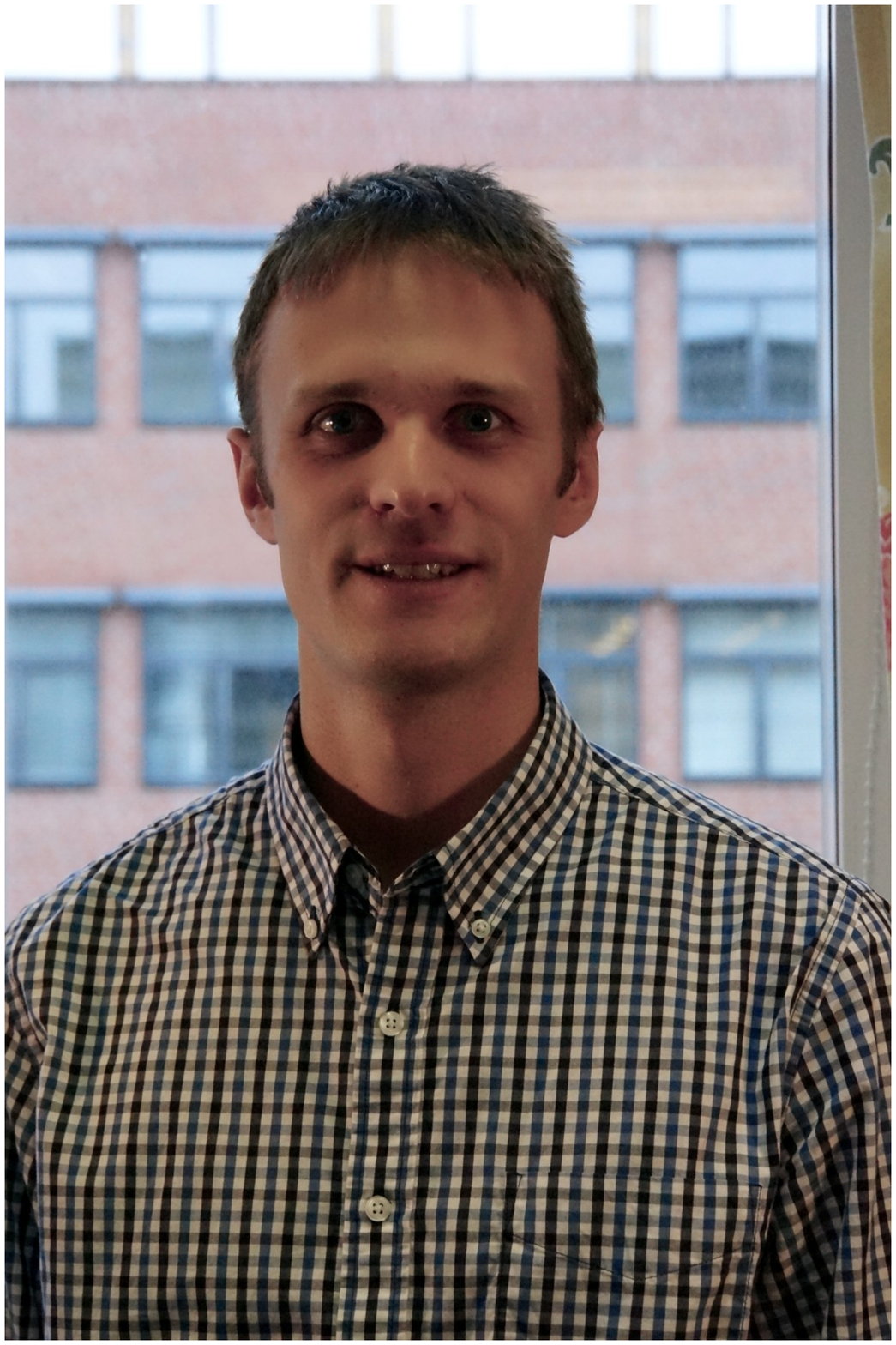}}]
{Anders Eklund}
(S'13) received the M.Sc. degree in engineering physics from KTH Royal Institute of Technology, Sweden, in 2011. He is currently working towards a Ph.D. degree in physics at KTH, experimentally investigating the frequency stability of spin torque oscillators by means of electrical characterization and synchrotron x-ray measurements.
\end{IEEEbiography}
\begin{IEEEbiography}[{\includegraphics[trim = 34mm 64mm 33mm 22mm width=1in,height=1.25in,clip,keepaspectratio]{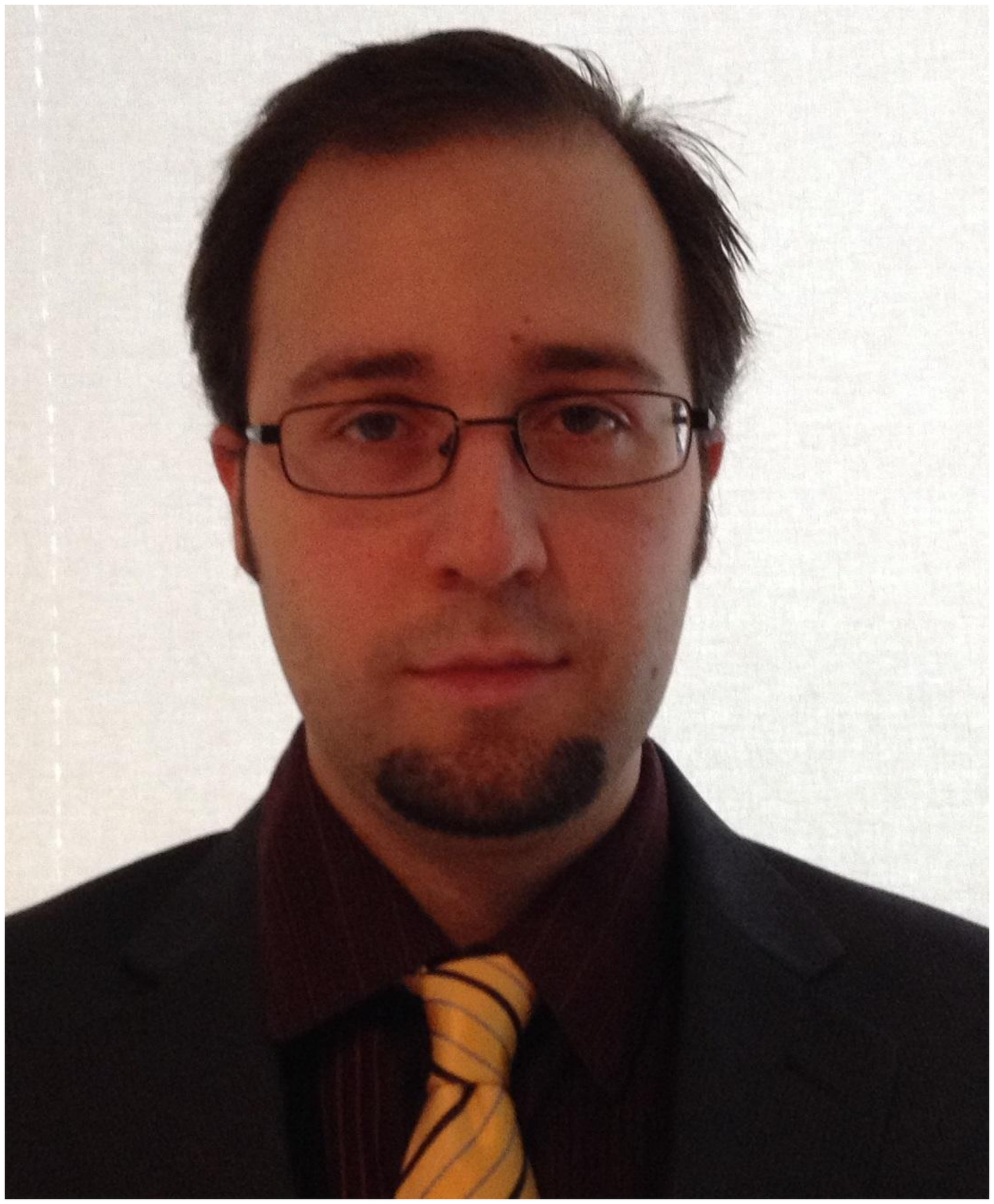}}]
{Ezio Iacocca}
(S'08) received the B.Sc. degree in electronic engineering from the Sim\'{o}n Bol\'{i}var University, Caracas, Venezuela ('08), the M.Sc. in nanotechnology from the Royal Institute of Technology, Stockholm, Sweden ('10), and the Ph.D. in physics from the University of Gothenburg, Gothenburg, Sweden ('14). His research focuses on the magnetodynamical modes of spin transfer torque driven nano oscillators and their applications in communication and storage technology.
\end{IEEEbiography}
\begin{IEEEbiography}[{\includegraphics[trim = 9mm 12mm 9mm 20mm width=1in,height=1.25in,clip,keepaspectratio]{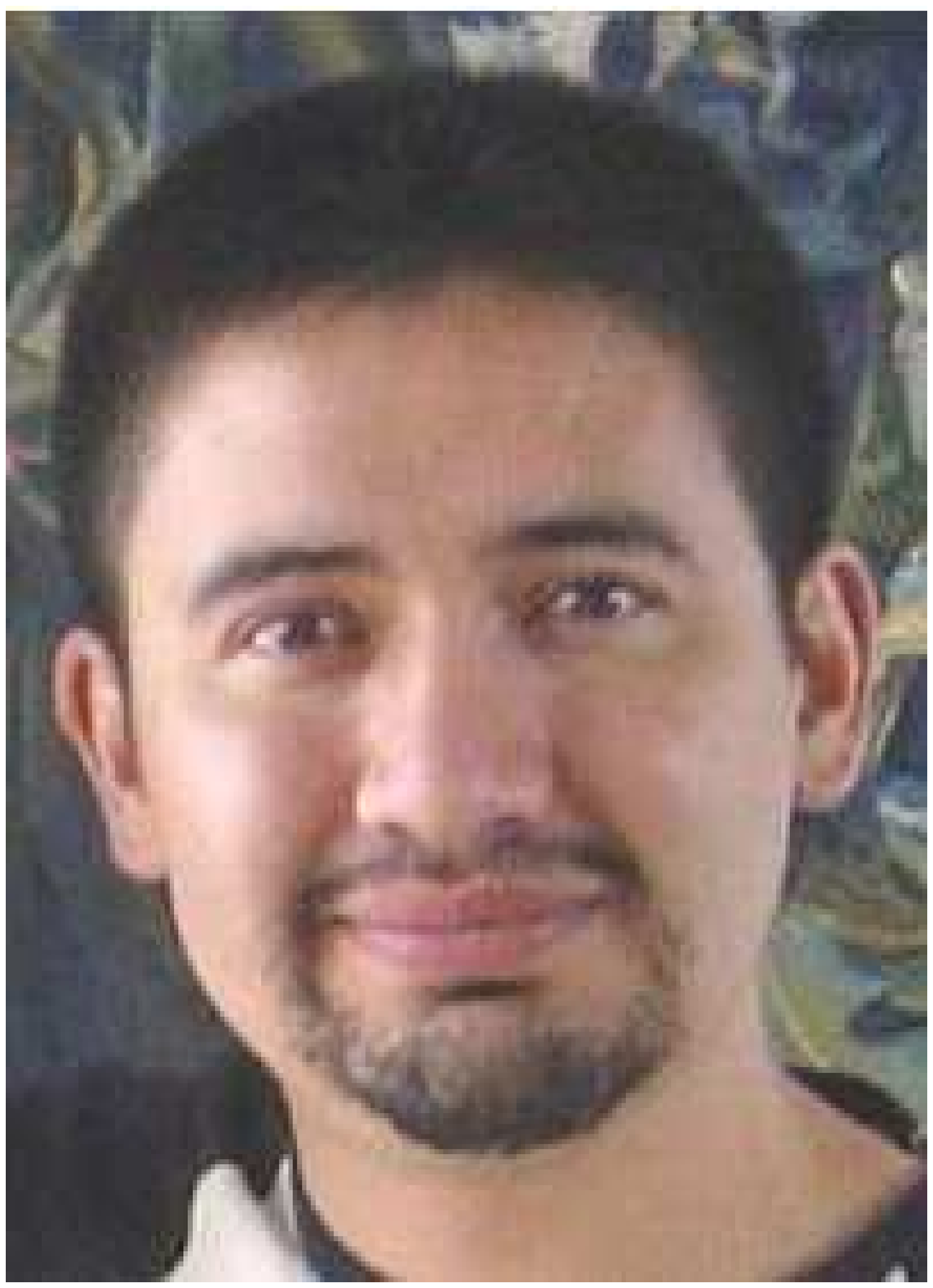}}]
{Saul Rodriguez}
(M'06) received the B.Sc. degree in electrical engineering from the Army Polytechnic School (ESPE), Quito, Ecuador, and the M.Sc. degree in system-on-chip design and the Ph.D. degree in electronic and computer systems from the KTH Royal Institute of Technology, Stockholm, Sweden. in 2001, 2005, and 2009, respectively. His current research interests include RF CMOS circuit design for wideband frond-ends, ultralow-power circuits for medical applications and graphene-based RF, and AMS circuits.
\end{IEEEbiography}
\begin{IEEEbiography}[{\includegraphics[trim = 80mm 10mm 80mm 19mm width=1in,height=1.25in,clip,keepaspectratio]{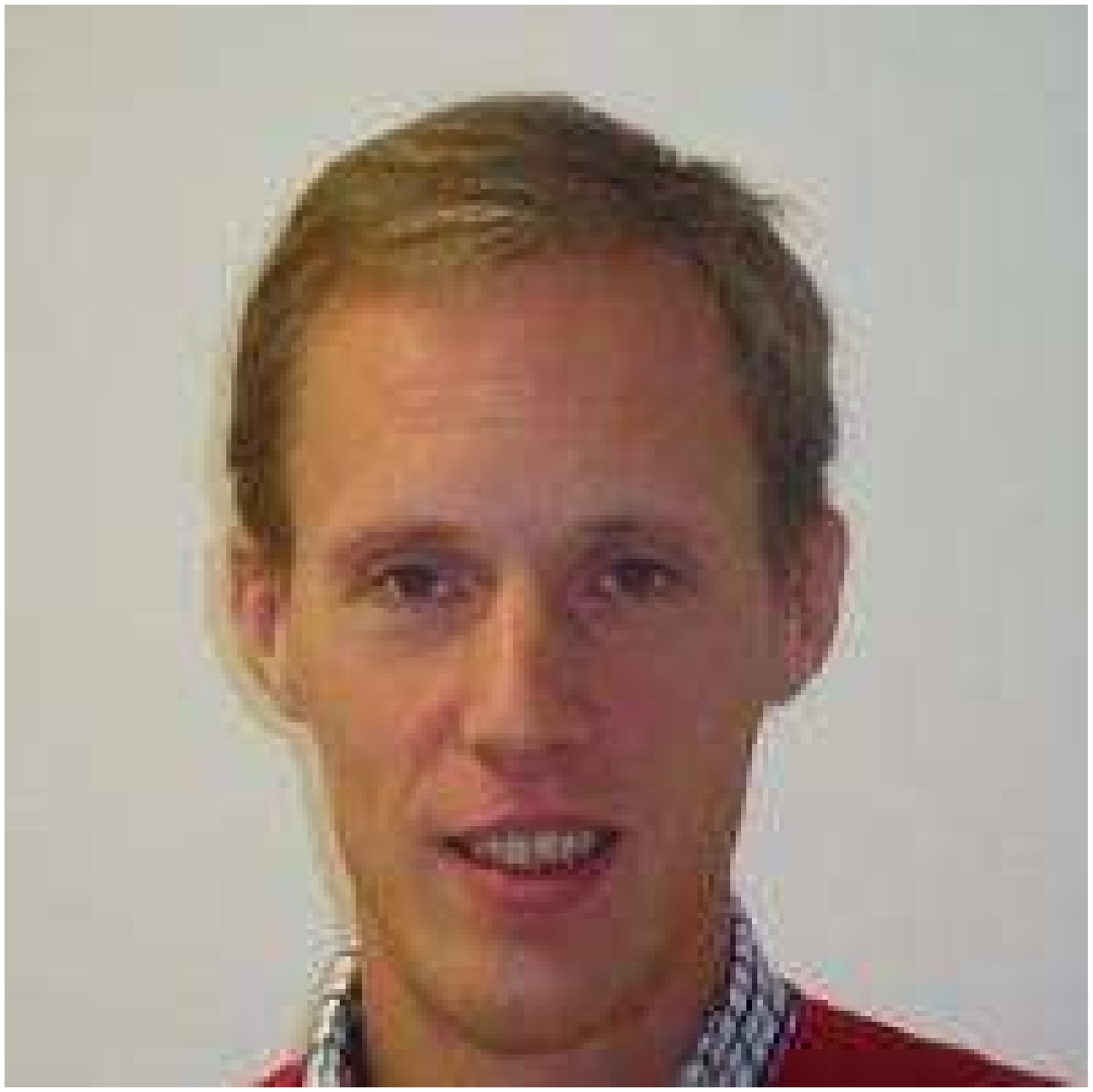}}]
{B. Gunnar Malm}
(M'98 - SM'10) was born in Stockholm, Sweden, in 1972. He received the M.S. from Uppsala University, Sweden, in 1997, the PhD in solid-state electronics 2002, from Royal Institute of Technology (KTH), Stockholm. He is an Associate Professor at the School of ICT, KTH since 2011. His recent work includes silicon photonics, silicon-carbide technology for extreme environments and spintronics. He also serves on the KTH Sustainability Council.
\end{IEEEbiography}
\begin{IEEEbiography}[{\includegraphics[trim = 57mm 44mm 120mm 10mm width=1in,height=1.25in,clip,keepaspectratio]{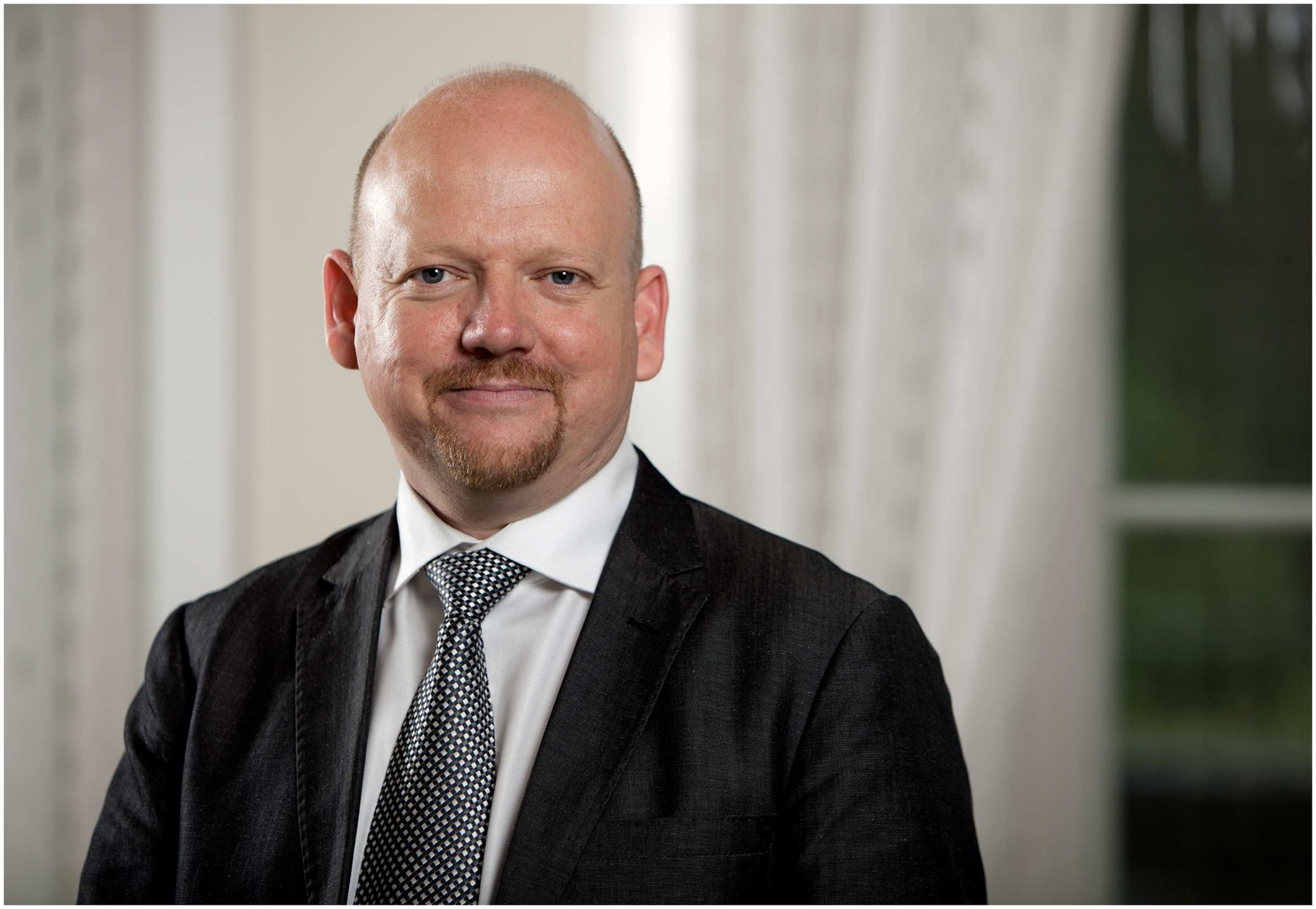}}]
{Johan $\AA$kerman}
(M'06) has an Ing. Phys. Dipl. degree ('94) from EPFL, Switzerland, a M.Sc. in physics ('96) from LTH, Sweden, and a Ph.D. in materials physics ('00) from KTH Royal Institute of Technology, Stockholm. In 2008 he was appointed Full Professor at University of Gothenburg and is a Guest Professor at KTH Royal Institute of Technology. He is also the founder of NanOsc AB and NanOsc Instruments AB.
\end{IEEEbiography}
\begin{IEEEbiography}[{\includegraphics[trim = 12mm 16mm 75mm 21mm width=1in,height=1.25in,clip,keepaspectratio]{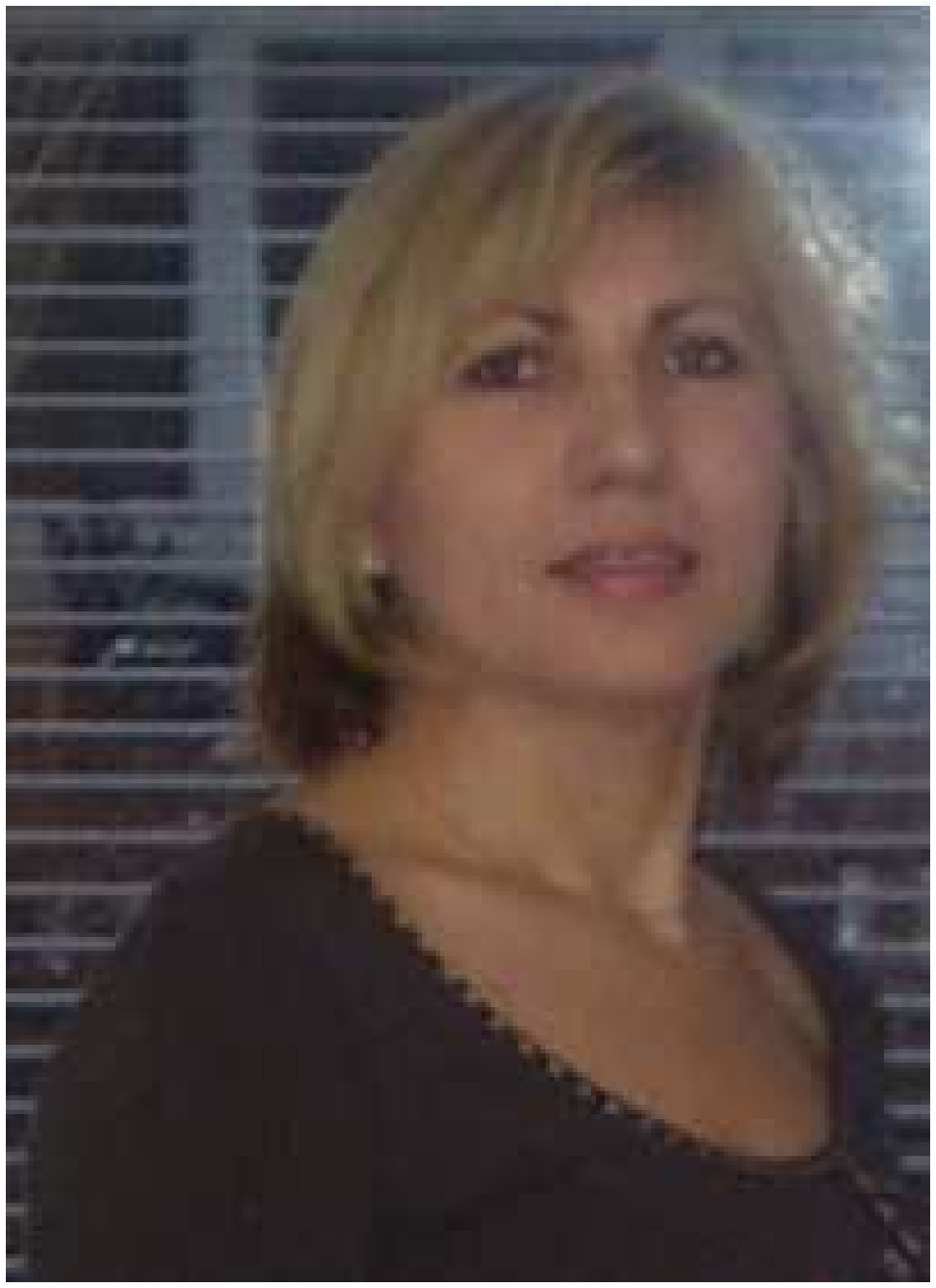}}]
{Ana Rusu}
(M'92) received the M.Sc. degree in electronics and telecommunications and Ph.D. degree in electronics in 1983 and 1998, respectively. She has been with KTH Royal Institute of Technology, Stockholm, Sweden, since 2001, where she is Professor in electronic circuits for integrated systems. Her research interests include low/ultralow power high performance CMOS circuits and systems, STO-based systems, RF graphene and high temperature SiC circuits.
\end{IEEEbiography}

\end{document}